\documentclass[aps, twocolumn, pra, showpacs]{revtex4}
\usepackage{graphicx}
\usepackage{amsmath}
\usepackage{amsfonts}
\usepackage{amssymb}
\usepackage{epsfig}

\begin{document}

\title{Density-functional fidelity approach to quantum phase transitions}
\author{Shi-Jian Gu}
\email{sjgu@phy.cuhk.edu.hk}
\affiliation{Department of Physics and ITP, The Chinese University of Hong Kong, Hong
Kong, China}

\begin{abstract}
We propose a new approach to quantum phase transitions in terms of the
density-functional fidelity, which measures the similarity between density
distributions of two ground states in parameter space. The key feature of
the approach, as we will show, is that the density-functional fidelity can
be measured easily in experiments. Both the validity and versatility of the
approach are checked by the Lipkin-Meshkov-Glick model and the
one-dimensional Hubbard model.
\end{abstract}

\pacs{64.60.-i, 71.15.Mb, 75.10.Jm, 71.10.Fd}
\date{\today}
\maketitle





\textit{Motivation}---Recently, considerable attentions \cite%
{HTQuan2006,Zanardi06,HQZhou0701,Buonsante1,WLYou07,PZanardi0701061,LCVenuti07,SJGu072,SChen07,MFYang07,HMPRE,XWang08032940,AHamma07,JHZhao0803,SYang08,DFAbasto08,XMLu032309,LCVenutiHub,LGong115114}
have been paid to the role of fidelity \cite{AUhlmann76,RJozsa94}, a concept
emerging from quantum information theory \cite{Nielsen1}, in quantum phase
transitions \cite{Sachdev}. Since the fidelity measures similarity of two
quantum states without considering any order information in the states, its
approach to quantum phase transitions has shed new lights on our understanding
of the quantum criticality. For example, the fidelity and its
leading term show singular and scaling behavior\cite%
{JHZhao0803,SYang08,AHamma07,DFAbasto08} in the topological phase transition%
\cite{wen-book} occurring in the ground states of both the Kitaev honeycomb
model \cite{Kitaevhoneycomb} and the Kitaev toric model \cite{Kitaevtoric}.
While the topoloigical phase transitions cannot be described by the
Landau-Ginzburg-Wilson spontaneous symmetry-breaking theory. Mathematically,
the fidelity is defined as an overlap between two ground states \cite%
{AUhlmann76,RJozsa94} separated by a certain distance in the parameter
space. To calculate the fidelity, one needs to know, in principle, all
information of the ground states, including their each component in the
Hilbert space of the Hamiltonian. This requirement is not an insuperable
barrier for theoretical studies because one can use some smart methods, for
instance, the density-matrix-renormalization-group technique \cite%
{MFYang07,SRWhiteDMRG,DMRGreview}, to reduce the dimension of the Hilbert
space. However, it might become a serious problem for experimental studies
because to obtain the full ground-state function of a quantum many-body
system is extremely difficult in experiments. This fact, to a certainty,
obstructs experimentalists from measuring the ground-state fidelity of
realistic quantum many-body systems.

Therefore, to find an appropriate approach that not only can describe the
change in the ground-state structure in perspective of information theory,
but also is based on experimentally measurable quantities, becomes a
nontrivial problem. In this paper, we propose a new approach to quantum
phase transitions in terms of the density-functional \cite%
{Hohenberg,KohnSham} fidelity (DFF) and its susceptibility (DFFS). According
to the Hohenberg-Kohn theorems \cite{Hohenberg}, the ground-state properties
of a quantum many-body system are uniquely determined by the density
distribution $n_{x}$ that minimizes the functional for the ground-state
energy $E_{0}[n_{x}]$. Therefore, the distribution $n_{x}$ captures the most
relevant information of the ground state. Any change in the structure of the
wavefunction can be found by calculating the similarity between two density
distributions, i.e. the DFF, directly. Since the density distribution is
usually measurable in experiments, our approach, therefore, provides a
practicable strategy to study quantum critical phenomena both theoretically
and experimentally. To check both the validity and versatility of the
approach, we take the Lipkin-Meshkov-Glick (LMG) model \cite{Lipkin1965} and
the one-dimensional Hubbard model \cite{Hubbard} as examples, and study the
DFF of real-space and momentum-space density distributions in the two
models, respectively. We show that the real-space DFF is able to witness the
second-order quantum phase transition \cite{HMPRE,CUT1,JVidalSpectra}
occurring in the ground state of the LMG model in the anisotropic case; and
the momentum-space DFF witness the Mott-insulator transition \cite{EHLieb},
which is of Beresinskii-Kosterlitz-Thouless (BKT) type \cite%
{VLBeresinskii,JMKosterlitz73}, in the ground state of the Hubbard model.

\textit{Formalism}---To begin with, we consider a general Hamiltonian of
quantum many-body systems%
\begin{equation}
\hat{H}(\lambda )=\hat{H}_{0}+\lambda \hat{H}_{I}+\sum_{x}\mu _{x}\hat{n}%
_{x},  \label{eq:Hamiltonian}
\end{equation}%
where $\hat{H}_{I}$ is the interaction term and $\lambda $ denotes strength,
and $\mu _{x}$ is the local (pseudo)potential associated with density
distribution $\{n_{x}\}$. The index $x$ can be discrete or continuous depending
on the system under study. Though in the local-density-approximation (LDA)
calculation, $n_{x}$ usually refers to the density of electrons in real space,
it can also be generalized to population in configuration space of a
reduced-density matrix or the density of state in energy(momentum) space. The
essence of the density-functional theory is that the density distribution has
already capture the most relevant information of the ground state, any change
in the environment conditions will leads another unique density distribution.
In quantum information perspective, the density-functional theory can be used
to calculate the ground-state entanglement in quantum many-body
systems\cite{LAWuDFT}.

Mathematically, the single-particle state can be obtained by tracing out all
other particles' degree of freedom, i.e. $\rho (x_{1},x_{1}^{\prime })=$tr$%
|\Psi _{0}(\lambda )\rangle \langle \Psi _{0}(\lambda )|$, where $|\Psi
_{0}(\lambda )\rangle $ is the ground state of the Hamiltonian (\ref%
{eq:Hamiltonian}). Then the density distribution $n$ is simply the diagonal
part of the reduced-density matrix $\rho (x_{1},x_{1}^{\prime })$%
\begin{equation}
n=\sum_{x}n_{x}|x\rangle \langle x|,\;\;\;\;\; n_x=\rho(x,x),
\label{eq:densitydis}
\end{equation}%
where $n$ has already been normalized. According to the Hellmann-Feymann
theorem, the density distribution can also be obtained as
\begin{equation}
n_{x}=\langle \Psi _{0}(\lambda )|\hat{n}_{x}|\Psi _{0}(\lambda )\rangle
=\langle \Psi _{0}(\lambda )|(\partial \hat{H}/\partial \mu _{x})|\Psi
_{0}(\lambda )\rangle .
\end{equation}%
Therefore, the change in the ground-state wavefunction can be reflected from
the fidelity between the two density distributions. For two ground states at
$\lambda $ and $\lambda ^{\prime }$, the DFF has the form,
\begin{equation}
F(\lambda ,\lambda^\prime )=\text{tr}\sqrt{n(\lambda )n(\lambda ^{\prime })}.
\end{equation}%
If we fix the distance $\delta \lambda =\lambda -\lambda ^{\prime }$, the DFF
is expected to show a drop around the critical point because two ground states
in different quantum phases has the maximum distance. In experiments, the
density distribution $n_{x}(\lambda )$ is measurable. Then the DFF between the
density distributions under different environmental condition can be obtained.

Meanwhile, it was found that the fidelity susceptibility
\cite{Zanardi06,WLYou07}, which denotes the leading term of the fidelity, plays
a central role in the fidelity approach to quantum phase transitions. Expanding
the DFF to the leading order, we can find that
\begin{equation}
F(\lambda ,\lambda +\delta \lambda )=1-\frac{\left( \delta \lambda \right)
^{2}}{2}\chi _{F}.
\end{equation}%
where the DFFS $\chi _{F}$ takes the form
\begin{equation}
\chi _{F}=\sum_{x}\frac{1}{4n_{x}}\left( \frac{\partial n_{x}}{\partial
\lambda }\right) ^{2}.  \label{eq:dfffs}
\end{equation}%
Therefore, if we regard $\partial n_{x}/\partial \lambda $ as an independent
function besides the density distribution $n_{x}$, \ the DFF is a functional of
$n_{x}$ and $\partial n_{x}/\partial \lambda $, both of which, in principle,
maximize the DFFS at the critical point. The typical case might be that the
denisty $n_{x}$ in a certain region $x_{\min }<x<x_{\max }$
\textit{vanishes}(for $1/n_{x}$) \textbf{rapidly}(for $\partial n_{x}/\partial
\lambda $\textbf{)}. These conclude the main formulism of the DFF approach to
quantum phase transitions.

\begin{figure}[tbp]
\includegraphics[width=8cm]{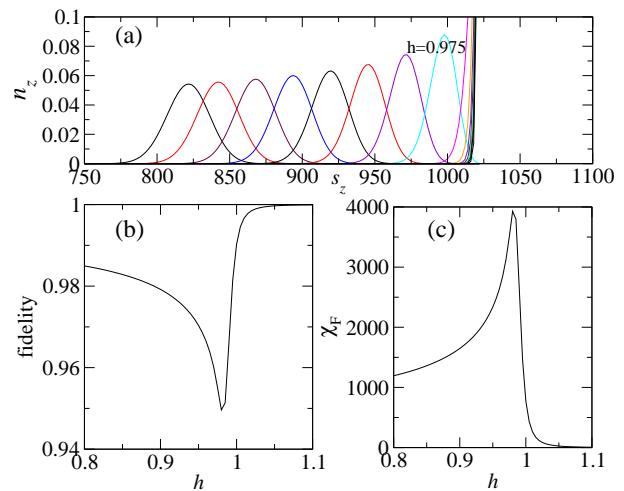}
\caption{(color online) The density-distribution properties of the LMG model
are analyzed for a sample of $S=1024$ and $\protect\gamma=0.5$. (a) The density
distribution $n_z$ as a function of $s_z$ for various $h=0.8, 0.825, \dots,
1.1$ (from left to right). The rightmost peak is
for $h=0.975$ (b) The DFF as a function of $h$ for $%
\protect\delta h=0.005$. (c) The DFFS as a function of $h$. }
\label{figure_lmgdff}
\end{figure}

\textit{The Lipkin-Meshkov-Glick (LMG) model}---To convince oneself of the
validity of the approach, let us first study the DFF in a simple quantum
phase transition occurring in the LMG model, which can provide us a clear
paradigm of the approach. The Hamiltonian of the LMG model can be written as
\begin{eqnarray}
H &=&-\frac{\lambda }{2S}\left( {1+\gamma }\right) \left( {\mathbf{\hat{S}}%
^{2}-\hat{S}_{z}^{2}-S}\right) -2h{\hat{S}}_{z}  \notag \\
&&-\frac{\lambda }{{4S}}\left( {1-\gamma }\right) \left( {\hat{S}_{+}^{2}+%
\hat{S}_{-}^{2}}\right) ,  \label{eq:HLMG}
\end{eqnarray}%
where ${\hat{S}}_{a}(a=x,y,z)$ is the spin-$S$ operator and ${\hat{S}}_{\pm
}={\hat{S}}_{x}\pm i{\hat{S}}_{y}$. The prefactor $1/S$ is necessary to
ensure a finite $E/S$ in the large $S$ limit. It is understood that the
total spin is the conserved quantities, i.e.,$\left[ {H,\mathbf{S}^{2}}%
\right] =0.$ The ground state of the Hamiltonian locates in the subspace of
maximum $S$ and can be expressed in the basis of ${\hat{S}}_{z}$, $|\Psi
_{0}(h)\rangle =\sum \varphi (s_{z})|s_{z}\rangle $. Therefore, the
ground-state energy is a density functional $E_{0}[n_{z}]$, where $%
n_{z}=\langle \Psi _{0}(h)|\hat{s}_{z}|\Psi _{0}(h)\rangle $ with $\hat{s}%
_{z}=|s_{z}\rangle \langle s_{z}|$.

The ground state of the LMG model in the anisotropic case ($\gamma \neq 1$)
consists of two phases, i.e the polarized phase for $h>1$, and the
symmetry-breaking phase for $0<h<1$. A simple quantum phase transition
occurs at the critical point $h_{c}=1$. For an illustrative \ purpose, we
show the density distributions $n_{z}$ as a function of $s_{z}$ for a sample
of $S=2048$ under various $h$s in Fig. \ref{figure_lmgdff} (a). When $h<1$,
the density distribution $n_{z}$ always shows a peak which is dragged to the
boundary of maximum $s_{z}$ little by little as $h$ increases. The DFF can
be sketched out from the overlap between two neighboring distributions,
which becomes smaller and smaller, and finally reaches a minimum around $%
h_{c}=1$. When $h>1$, the density peak diverges. In this case, a small
change of $h$ can not change the density distribution significantly, then
the DFF is close to 1 again. Fig. \ref{figure_lmgdff} (b) represents the DFF
as a function of $h$ for $\delta h=0.005$. Clearly the sudden drop of the
DFF corresponds the second-order quantum phase transition. The sharp peak in
the DFF susceptibility [Fig. \ref{figure_lmgdff} (c)] implies that the
density distribution evolves dramatically around the critical point.

For the LMG model, $n_{z}=\varphi ^{\ast }(s_{z})\varphi (s_{z})$, so the
DFF is mathematically the same as the previously studied ground-state
fidelity \cite{HMPRE}, hence satisfies the same scaling and critical
behaviors as the ground-state fidelity. However, we would like to emphsis
here that, the motivation of the DFF is quite different. The mathematical
coincidence of two fidelities of the LMG model is due to that the model in
the subspace of maximum $S$ becomes a single-particle problem. In this case,
the diagonal element of the single-particle reduced-density matrix is just
the absolute value of the wavefunction.

\textit{The one-dimensional Hubbard model}--- The density distribution in the
density-functional theory can be generalized to any density distribution, for
instance, the density of state. To see the versatility of the approach, here we
take the one-dimensional Hubbard model \cite{Hubbard,EHLieb} to show how the
change in the density of state in momentum space around the critical point can
be reflected from the corresponding DFF and DFFS. The Hamiltonian of the
one-dimensional Hubbard model reads
\begin{equation}
H=-\sum_{\sigma ,j=1}^{L}(\hat{c}_{j,\sigma }^{\dagger }\hat{c}_{j+1,\sigma
}+\hat{c}_{j+1,\sigma }^{\dagger }\hat{c}_{j,\sigma })+U\sum_{j}\hat{n}%
_{j,\uparrow }\hat{n}_{j,\downarrow }
\end{equation}%
where $\hat{c}_{j,\sigma }^{\dagger }$ and $\hat{c}_{j,\sigma },\sigma
=\uparrow ,\downarrow $ are creation and annihilation operators for
fermionic atoms with spin $\sigma $ at site $j$ respectively, $\hat{n}%
_{\sigma }=\hat{c}_{\sigma }^{\dagger }\hat{c}_{\sigma }$, and $U$ denotes the
strength of on-site interaction. The Hubbard model can be solved exactly via
the Bethe-ansatz method \cite{EHLieb}. For the case of $U>0$ and $N\leq L $,
the energy spectra of the system under the periodic boundary conditions are
determined by a set of charge and spin rapidities $\{k_{j},\lambda _{a}\} $,
which satisfy the following transcendental Bethe-ansatz equations(BAEs)
\begin{eqnarray}
&&2\pi I_{j}=k_{j}L-\sum_{a=1}^{M}\theta _{1}(\lambda _{a}-\sin k_{j}),
\label{eq:BAE1} \\
&&2\pi J_{a}=\sum_{j=1}^{N}\theta _{1}(\lambda _{a}-\sin
k_{j})-\sum_{b=1}^{M}\theta _{2}(\lambda _{a}-\lambda _{b}),  \label{eq:BAE2}
\end{eqnarray}%
where $\theta _{n}(k)=2\tan ^{-1}(4k/nU)$, $M$ is the number of down spins
and $\{I_{j},J_{a}\}$ play the role of quantum number. The ground-state
solution only consists of real $k$s and real $\lambda $s. It is a singlet
state, given by a quantum number configuration (successive integers or
half-odd-integers) symmetrically arranged around zero. Once the rapidities $%
\{k_{j},\lambda _{a}\}$ are obtained from Eqs. (\ref{eq:BAE1},\ref{eq:BAE2}%
), the energy of the state can be calculated as $E=-2\sum_{j}\cos k_{j}$.

For the half-filled case, i.e $N=L$, a Mott-insulator transition occurs at the
critical point $U=0$. If $U>0$, a charge gap opens and the system becomes an
insulator. The ground-state energy the Hubbard model around $U=0$ is infinitely
differentiable, the phase transition belongs to the BKT type. Recent
entanglement approach to the Hubbard model \cite{SJGuPRL,DLarsson05} shows that
single-site entanglement shows a maximum, due to the various correlations have
long-range behaviors, at the critical point. However, whether the ground-state
fidelity can witness the Mott-insulator transition is still controversial
\cite{WLYou07,LCVenutiHub}. To apply the DFF approach to the phase transition,
we use the concept of the density of state in the quasi-momentum space that is
defined as $\rho \lbrack (k_{j+1}+k_{j})/2]=1/[L(k_{j+1}-k_{j})]$. Then the
ground-state energy can be calculated as $E=-(N/\pi )\int_{-\pi }^{\pi }dk\rho
(k)\cos k$. Clearly, the density of state $\rho (k)$ uniquely decides the
ground-state energy.

\begin{figure}[tbp]
\includegraphics[width=8cm]{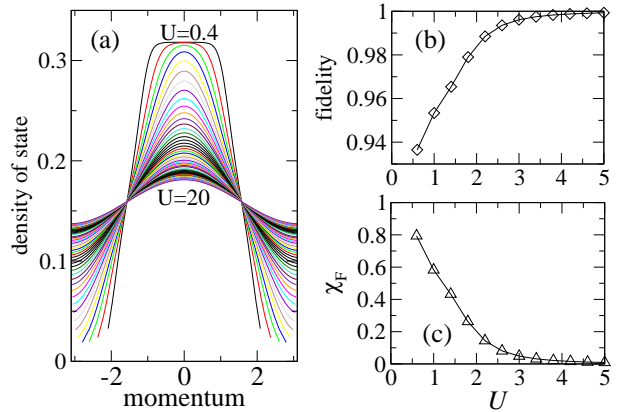}
\caption{(color online) The density of state and DFF of the Hubbard model are
analyzed for a finite sample of $L=N=210$. (a) The density of state for various
on-site interaction $U=20, 19.6,\dots, 0.04$ with step 0.04. Two ends of each
line denotes the momentum cut-off in finite systems. (b)The DFF as a function
of $U$. (c) The DFFS a function of $U $.} \label{figure_hugdff}
\end{figure}

To see the evolution in the density of state, in Fig. \ref{figure_hugdff}
(a), we show the density of state for a system of $L=N=210$ and a set of
uniformly distributed $U$ with space $\delta U=0.04$ . From the figure, we
can see that if $U$ is large, the densities of state are very close, this
observation means that the DFF between two closing densities is almost 1
[See Fig. \ref{figure_hugdff} (b)]. The underlying physical picture is that,
in the large $U$ limit, the system becomes a full-filled spinless fermion
model, the $k$s in Eq. (\ref{eq:BAE1}) are almost uniformly distributed,
then the density of state is very flat in the momentum space. On the other
hand, as $U$ becomes smaller and smaller, the second term of Eq. (\ref%
{eq:BAE1}) makes any $k$ more close to the zero point hence suppresses the
density of state in the region of $|k|>\pi /2$, while increase $\rho (k)$ in
$|k|<\pi /2$. The physics is straightforward. The Pauli's exclusion principle
does not forbidden two electrons with different spin polarizations to occupy
the same state in $k$ space. To have a low energy, electrons want to stay as
close as possible to the $k=0$ point. Then the density of state become higher
in $|k|<\pi /2$ and smaller in $|k|>\pi /2$ as $U$ deceases. Around the
critical point, the change in the density of state becomes dramatic. Then the
DFF deviates from 1 little by little, and the fidelity susceptibility becomes
larger and larger. Though the Bethe-ansatz equations are extremely difficult to
be solved around $U=0,$ we can still find the DFFS tends to a maximum as $U$
tends to zero.

We can also check the above picture analytically in some special cases. In
the thermodynamic limit and half-filled case, the density of state is \cite%
{EHLieb}
\begin{equation}
\rho (k)=\frac{1}{2\pi }+\frac{\cos k}{\pi }\int_{0}^{\infty }\frac{%
dpJ_{0}(p)\cos (p\sin k)}{1+e^{U|p|/8}},  \label{eq:hubdos}
\end{equation}
where $J_0$ is zeroth order Bessel function. If $U=\infty $, the second term of
Eq. (\ref{eq:hubdos}) vanishes, so $\rho
(k)=1/2\pi $ for $k\in \lbrack -\pi ,\pi ]$. If $U=0$, since $%
\int_{0}^{\infty }dpJ_{0}(p)\cos (p\sin k)=|\cos k|^{-1},$ so $\rho (k)=1/\pi $
for $k\in \lbrack -\pi /2,\pi /2]$ and $\rho (k)=0$ for $|k|>\pi /2$. This fact
tells us that the density of state undergoes a dramatic
change around the critical point. Therefore, as implied from the Eq. (\ref%
{eq:dfffs}), the DFF should reaches a maximum at the critical point $U=0$.

In summary, we have proposed a new approach to quantum phase transitions based
on the DFF and DFFS. The former measures the similarity between two density
distributions and the latter describes the changing rate of the density
distribution in the parameter space. We show that the divergence of the DFFS is
typically related to a quickly vanishing density distribution, which actually
denotes a quantum phase transition. To check the validity of the approach, we
studied the DFF in the LMG model and show that the divergence of the DFF at the
critical point. Also to see the versatility of the approach, we studied the DFF
for the density of state of the one-dimensional Hubbard model. As the on-site
$U$ tends to zero, the density of state in the regions $[-\pi ,-\pi /2]$ and
$[\pi /2,\pi ]$ quickly vanishes. This phenomena leads a maximized (at least)
DFFS around the critical point $U=0$.

Though we restricted out studies in two well-studied strongly correlated
systems, the approach we proposed can be applied to any quantum systems.
Especially, the DFF approach can be used to study various quantum phase
transitions (like the structural phase transition) based the LDA calculations.

We thank H. Q. Lin and X. Wang for helpful discussions. This work is
supported by the Direct grant of CUHK (A/C 2060344).

\end{document}